\documentclass[conference]{IEEEtran}
\usepackage{orcidlink}
\usepackage{cite}
\usepackage{amsmath,amssymb,amsfonts}
\usepackage{algorithmic}
\usepackage{algorithm}
\usepackage{graphicx}
\DeclareMathOperator*{\argmin}{arg\,min}
\usepackage{textcomp}
\usepackage{xcolor}
\usepackage{amsmath}
\usepackage{subcaption}
\makeatletter

\def\ps@IEEEtitlepagestyle{%
  \def\@oddfoot{\mycopyrightnotice}%
  \def\@evenfoot{}%
}
\def\mycopyrightnotice{%
  \gdef\mycopyrightnotice{}
}

\usepackage{blindtext}
\usepackage{eso-pic}
\IEEEoverridecommandlockouts
\usepackage{cite}
\usepackage{amsmath,amssymb,amsfonts}
\usepackage{algorithmic}
\usepackage{graphicx}
\usepackage{textcomp}
\usepackage{xcolor}
\def\BibTeX{{\rm B\kern-.05em{\sc i\kern-.025em b}\kern-.08em
    T\kern-.1667em\lower.7ex\hbox{E}\kern-.125emX}}
    
\usepackage{eso-pic}

\begin{document}
\title{\vspace*{1cm} Deep Reinforcement Learning with Local Interpretability for Transparent Microgrid Resilience Energy Management\\

}

\author{\IEEEauthorblockN{Mohammad Hossein \\Nejati Amiri}
\IEEEauthorblockA{\textit{College of Engineering} \\
\textit{Birmingham City University}\\
Birmingham, UK \\
mohammadhossein.nejatiamiri\\@mail.bcu.ac.uk \orcidlink{0000-0003-0195-4087}
}
\and
\IEEEauthorblockN{Fawaz Annaz}
\IEEEauthorblockA{\textit{College of Engineering} \\
\textit{Birmingham City University}\\
Birmingham, UK \\
fawaz.annaz@bcu.ac.uk \orcidlink{0000-0002-9236-9387}}
\and
\IEEEauthorblockN{Mario De Oliveira}
\IEEEauthorblockA{\textit{College of Engineering} \\
\textit{Birmingham City University}\\
Birmingham, UK \\
mario.deoliveira@bcu.ac.uk \orcidlink{0000-0002-3619-3989}}
\and
\IEEEauthorblockN{Florimond Gueniat*}
\IEEEauthorblockA{\textit{College of Engineering} \\
\textit{Birmingham City University}\\
Birmingham, UK \\
florimond.gueniat@bcu.ac.uk \orcidlink{0000-0001-6035-2540}}
}

\maketitle
\begin{abstract}
Renewable energy integration into microgrids has become a key approach to addressing global energy issues such as climate change and resource scarcity.
However, the variability of renewable sources and the rising occurrence of High Impact Low Probability (HILP) events require innovative strategies for reliable and resilient energy management.
This study introduces a practical approach to managing microgrid resilience through Explainable Deep Reinforcement Learning (XDRL).
It combines the Proximal Policy Optimization (PPO) algorithm for decision-making with the Local Interpretable Model-agnostic Explanations (LIME) method to improve the transparency of the actor network's decisions.
A case study in Ongole, India, examines a microgrid with wind, solar, and battery components to validate the proposed approach. 
The microgrid is simulated under extreme weather conditions during the Layla cyclone. LIME is used to analyse scenarios, showing the impact of key factors such as renewable generation, state of charge, and load prioritization on decision-making.
The results demonstrate a Resilience Index (RI) of 0.9736 and an estimated battery lifespan of 15.11 years. 
LIME analysis reveals the rationale behind the agent's actions in idle, charging, and discharging modes, with renewable generation identified as the most influential feature.
This study shows the effectiveness of integrating advanced DRL algorithms with interpretable AI techniques to achieve reliable and transparent energy management in microgrids.
\end{abstract}



\section{Introduction}

The electricity grid has witnessed major changes in recent years, driven by advancements such as bidirectional communication between suppliers and consumers, enabling smart grids' development.
Smart grids incorporate concepts like microgrids, demand response, prosumers, self-healing systems, and local electricity markets to meet evolving energy needs \cite{nejati2023heuristic}.
Renewable energy sources have received significant attention as a response to global warming, the push for net-zero emissions, and the challenges of fossil fuel depletion and pollution \cite{han2024driving}.
Microgrids, with their ability to function in both isolated and grid-connected modes, serve as a practical framework for leveraging renewable energy despite its intermittent nature \cite{amiri2024strategies}.

Modern power grids face growing vulnerabilities from natural disasters and cyber threats, intensified by digitalisation.
These rare but high-impact events underscore the importance of enhancing resilience in smart grids \cite{amiri2024towards}.
At the same time, advances in Artificial Intelligence (AI) have provided new tools to tackle challenges in power systems, such as real-time operations and uncertainties like variable loads (e.g., electric vehicles) and energy production \cite{arevalo2024impact}.
Reinforcement learning (RL) stands out among AI approaches for its ability to quickly adapt to dynamic environments and effectively manage systems without the need for detailed models.
This makes RL particularly suitable for addressing various planning and operational challenges in power systems, such as optimal dispatch and control \cite{mohammadi2024comparative}.

Recent advancements in Deep Reinforcement Learning (DRL), which combines RL with Deep Neural Networks (DNN), have shown outstanding performance in solving complex problems.
Despite its potential, DRL's application in critical industries such as power systems is still limited due to concerns over its black-box nature and the lack of transparency in decision-making \cite{stavrev2024reinforcement}.
To tackle concerns about the black-box nature of DNN, along with ethical issues and regulations like the General Data Protection Regulation (GDPR), eXplainable Artificial Intelligence (XAI) has become a prominent research area in computer science \cite{chamola2023review}.
In DRL, eXplainable Deep Reinforcement Learning (XDRL) is classified into three approaches: Interpretable Agents (IA), Intrinsic Explainability (IE), and Post-hoc Explainability (PHE) \cite{bekkemoen2024explainable}.

Interpretable Agents (IA) are inherently designed to be easily understood by humans, relying on rule-based or linear models.
However, this simplicity often comes at the cost of performance.
Intrinsic Explainability (IE) focuses on enhancing explainability by modifying the RL agent or its model, such as reward and transition functions, as part of its design. 
Post-hoc Explainability (PHE) generates explanations for the decisions of pre-trained models through external techniques without altering their internal structure.

Among these approaches, PHE offers high performance despite its limited inherent explainability, which makes it well-suited for smart grid applications where performance is a priority.
Since it operates after model training, PHE can also be used to analyse pre-existing models.
Methods like Local Interpretable Model-agnostic Explanations (LIME) and SHapley Additive exPlanations (SHAP) are commonly used, with LIME offering local insights and SHAP providing both local and global explanations \cite{LIME_ORIG,SHAP_ORIG}.
This study uses LIME for its simplicity and focus on specific decision analysis.

The main contributions of this work are as follows:

\begin{itemize}
    \item A real-world case study in Ongole (India), utilizing actual geographical and load data for microgrid design. 
    The microgrid comprises wind turbines, solar panels, batteries, and loads with different priority levels.
    Component sizing is performed using HOMER Pro to achieve an optimal design tailored to the location's specific needs.
    \item Renewable generation simulated under the impact of the Layla cyclone that occurred in this region in mid-May 2010 to capture the challenges posed by extreme weather events.
    \item Use of the DRL Proximal Policy Optimization (PPO) algorithm within an Actor-Critic framework, a well-established and efficient method, enhances the resilience of energy management of the microgrid.
    \item Application of the LIME method to explain specific decisions made by the DRL agent. 
    This strategy enhanced transparency and built stakeholder trust in the system's operations.
\end{itemize}
By integrating these features, the study aims to present an explainable approach to microgrid resilience management that aligns with real-world scenarios and stakeholder expectations.

The structure of this article is as follows:
Section \ref{section:MG_Modell} describes the microgrid modelling approach, the LIME method, and the mathematical formulation used in the study.
Section \ref{section:Sim_Result} presents and analyses the simulation results.
Finally, Section \ref{section:Conclusion} concludes the study and outlines potential future research directions.

\section{Microgrid Modelling} \label{section:MG_Modell}
\begin{figure*}[!]
    \centering
    \includegraphics[width=0.67\textwidth]{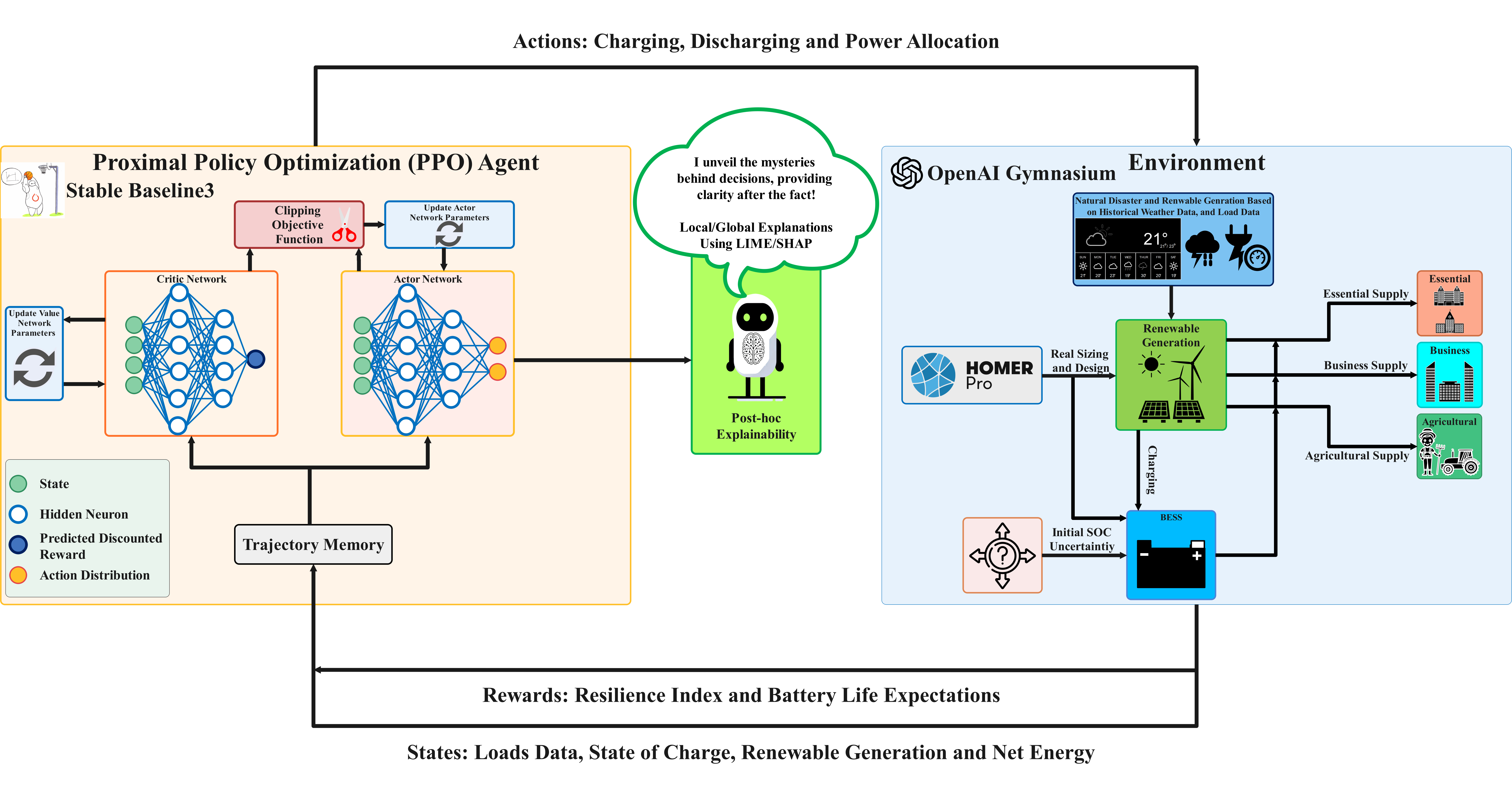} 
    \caption{Proposed XDRL Framework for Microgrid Resilience Energy Management.}
    \label{fig:XDRL_Model}
\end{figure*}

\begin{figure}[!]
    \centering
    \includegraphics[width=0.48\textwidth]{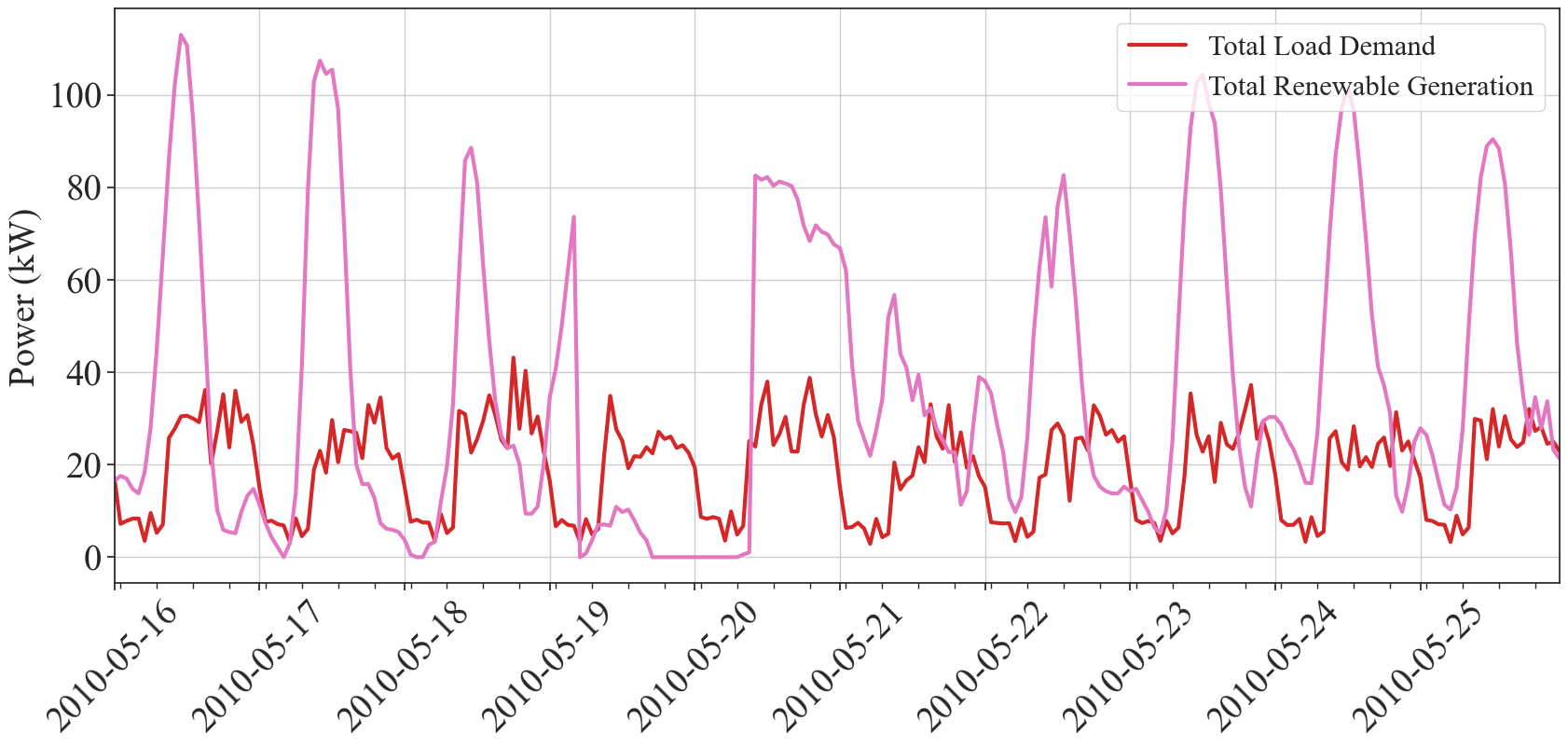}
    \caption{Total Load and Total Renewable Power Generation Profiles Based on Weather Data.}
    \label{fig:Total_load_gen}
\end{figure}

We used HOMER Pro to size an appropriate microgrid for the considered site, consisting of 140 kW of solar power, 80 kW of wind power, a 780 kWh battery, and a 52 kW converter. 
Figure \ref{fig:XDRL_Model} illustrates the resilient energy management of the microgrid using XDRL.
The total load consumption and the renewable power generated based on the weather data are depicted in Figure \ref{fig:Total_load_gen}.


The following section discusses modelling the RL environment.

\subsection{Microgrid Environment Design}

The design of the microgrid environment is crucial for the DRL agent to interact effectively and learn optimal policies. 
The environment encapsulates the physical and operational characteristics of a microgrid, including battery storage, renewable energy generation, and load demands.
The custom environment is developed using the OpenAI Gymnasium framework, which offers a standardised interface for DRL agents. 
It allows libraries like Stable-Baselines3 to seamlessly interact with the environment.
The environment simulates battery operations, renewable energy inputs, and load demands over discrete time steps.
Renewable energy generation and load demands are deterministic, based on historical data, while the initial State Of Charge (SOC) is randomized to introduce variability in initial conditions.

The state space represents all possible states of the microgrid at any given time step. 
It includes variables essential for the agent to make informed decisions. 
The state vector at time step \( t \) is defined as:

\begin{equation} 
\mathbf{s}_t = \left[ \text{SOC}_t, L_{1,t}, L_{2,t}, L_{3,t}, P_{\text{RE},t}, P_{\text{net},t} \right].
\end{equation}

The SOC of the battery at time \( t \), denoted as \(\text{SOC}_t\), is constrained between \(\text{SOC}_{\min} = 0.2\) and \(\text{SOC}_{\max} = 0.9\) to consider the safety and longevity of the battery.
The load demand at time \( t \) is represented by \( L_{i,t} \), where \( i = 1, 2, 3 \) corresponds to different priority levels.
Essential load (\( L_{1,t} \)) holds the highest priority with a weight 3.5 times that of business load (\( L_{2,t} \)), which itself has 2 times higher priority than agricultural load (\( L_{3,t} \)), the lowest priority. 

Renewable energy generation (\( P_{\text{RE},t} \)) at time \( t \) is determined from historical weather data, and the net energy (\( P_{\text{net},t} \)) is calculated as: 

\begin{equation} 
P_{\text{net},t} = P_{\text{RE},t} - \sum_{i=1}^{3} L_{i,t}.
\end{equation}

The action space defines the set of all possible actions that the agent can take at each time step.
It is a continuous space consisting of five variables, represented as:

\begin{equation}  
\mathbf{a}_t = \left[ a_{\text{ch},t}, a_{\text{dis},t}, w_{1,t}, w_{2,t}, w_{3,t} \right],
\end{equation}
where \( a_{\text{ch},t} \) and \( a_{\text{dis},t} \) are the normalized charging and discharging power actions, respectively, both constrained to the range \([-1, 1]\). 
The variables \( w_{i,t} \), for \( i = 1, 2, 3 \), represent the raw weights for distributing power among the three loads, and each is also bounded within \([-1, 1]\). 
This action space enables the agent to determine charging and discharging actions while allocating power to different loads effectively.





The weights for power distribution among the loads are normalised using a softmax function to ensure they sum to one:

\begin{equation}
\hat{w}_{i,t} = \frac{\exp(w_{i,t})}{\sum_{j=1}^{3} \exp(w_{j,t})}, \quad \text{for } i = 1, 2, 3.
\end{equation}

The SOC of the battery is updated based on charging and discharging actions while considering efficiency losses:

\begin{equation}
\text{SOC}_{t+1} = \text{SOC}_t + \frac{\eta_{\text{ch}} P_{\text{ch},t} - \frac{P_{\text{dis},t}}{\eta_{\text{dis}}}}{E_{\text{max}}},
\end{equation}

subject to the constraint:

\begin{equation}
\text{SOC}_{\min} \leq \text{SOC}_{t+1} \leq \text{SOC}_{\max},
\end{equation}
where \(\eta_{\text{ch}} = 0.90\) is the charging efficiency, \(\eta_{\text{dis}} = 0.95\) is the discharging efficiency, and \(E_{\text{max}} = 780 \, \text{kWh}\) is the maximum battery capacity. The available capacities for charging and discharging are defined as:

\begin{equation}
E_{\text{avail,ch},t} = (\text{SOC}_{\max} - \text{SOC}_t) E_{\text{max}},
\end{equation}
\begin{equation}
E_{\text{avail,dis},t} = (\text{SOC}_t - \text{SOC}_{\min}) E_{\text{max}}.
\end{equation}

The charging and discharging powers are further constrained by the available capacities:

\begin{equation}
P_{\text{ch},t} = \min\left( P_{\text{ch},t}, E_{\text{avail,ch},t} \right),
\end{equation}
\begin{equation}
P_{\text{dis},t} = \min\left( P_{\text{dis},t}, E_{\text{avail,dis},t} \right).
\end{equation}

Additionally, charging and discharging cannot occur simultaneously. Depending on the net energy, either charging or discharging is permitted:

\begin{equation}
\begin{cases} 
P_{\text{dis},t} = 0, & \text{if } P_{\text{net},t} \geq 0, \\
P_{\text{ch},t} = 0, & \text{if } P_{\text{net},t} < 0.
\end{cases}
\end{equation}

The net power supply after battery operation (\( P_{s,t} \)) is calculated as:

\begin{equation}
P_{s,t} = P_{\text{RE},t} + P_{\text{dis},t} - P_{\text{ch},t}.
\end{equation}

This power is allocated to the loads based on the normalized weights:

\begin{equation}
P_{s,i,t} = \hat{w}_{i,t} P_{s,t}, \quad \text{for } i = 1, 2, 3.
\end{equation}

The power imbalance for each load is given by:

\begin{equation}
P_{\text{imb},i,t} = P_{s,i,t} - L_{i,t}, \quad \text{for } i = 1, 2, 3.
\end{equation}

Negative imbalances indicate shortages, which are critical for resilience calculations. These shortages are defined as:

\begin{equation}
P_{\text{sh},i,t} = -\min(0, P_{\text{imb},i,t}), \quad \text{for } i = 1, 2, 3.
\end{equation}

The reward function guides the agent towards actions that enhance system resilience.
At each time step, the reward \( r_t \) calculates the Resiliency Index (RI) reward.

The RI reward incentivizes the agent to minimize power shortages, especially for high-priority loads. 
It is calculated as:
\begin{equation}
r_{t} = \left( 1 - \frac{7 P_{\text{sh},1,t} + 2 P_{\text{sh},2,t} + 1 P_{\text{sh},3,t}}{7 L_{1,t} + 2 L_{2,t} + 1 L_{3,t}} \right).
\end{equation}

The cumulative reward for the entire episode is obtained by summing over all time steps \( t \) from the start to the termination of the episode:

\begin{equation} 
R_{\text{episode}} = \sum_{t=1}^{T} r_t.
\end{equation}

The overall RI for the episode is calculated as:

\begin{equation}
\resizebox{0.435\textwidth}{!}{$
\text{RI} = 1 - \frac{7 \sum_{t=1}^{T} P_{\text{sh},1,t} + 2 \sum_{t=1}^{T} P_{\text{sh},2,t} + 1 \sum_{t=1}^{T} P_{\text{sh},3,t}}{7 \sum_{t=1}^{T} L_{1,t} + 2 \sum_{t=1}^{T} L_{2,t} + 1 \sum_{t=1}^{T} L_{3,t}}.
$}
\end{equation}

The total reward for the episode is then adjusted by adding the contributions from the overall RI at the termination of the episode:

\begin{equation} 
R_{\text{final}} = R_{\text{episode}} + \text{RI}.
\end{equation}

Finally, the total reward for the episode is normalized by dividing it by the maximum possible total reward \( r_{\text{max}} \):

\begin{equation} 
R_{\text{final}}^{\text{normalized}} = \frac{R_{\text{final}}}{r_{\text{max}}}.
\end{equation}

\subsection{Conceptual Overview: PPO and LIME}

PPO is a DRL algorithm designed to improve policy-gradient methods by ensuring stable and efficient training. 
PPO simplifies Trust Region Policy Optimization (TRPO) by avoiding complex second-order optimization while maintaining stable updates.
The core idea of PPO is to use a clipped surrogate objective that limits excessively large policy updates \cite{PPO_ORIG}.
The policy update is formulated as:  

\begin{equation}
\resizebox{0.435\textwidth}{!}{$
L^{\text{CLIP}}(\theta) = \mathbb{E}_t \big[ \min \big( r_t(\theta) \hat{A}_t, 
\text{clip}(r_t(\theta), 1 - \epsilon, 1 + \epsilon) \hat{A}_t \big) \big],
$}
\end{equation}
where \( r_t(\theta) = \frac{\pi_\theta(a_t | s_t)}{\pi_{\theta_{\text{old}}}(a_t | s_t)} \) is the probability ratio between the new policy \(\pi_\theta\) and the old policy \(\pi_{\theta_{\text{old}}}\), \(\hat{A}_t\) is the advantage function, and \(\epsilon\) is the clipping parameter (e.g., 0.2). 
The agent follows a policy \(\pi_\theta\) that determines its actions to optimize rewards over time, with identifying the policy equivalent to determining its parameters \(\theta\). 
This objective ensures that policy updates remain within a trust region, avoiding performance degradation from overly large updates.


The PPO algorithm alternates between collecting experience through interaction with the environment and optimizing the policy using gradient ascent. The combined objective includes terms for the value function \( L^{\text{VF}} \) and an entropy bonus \( L^{\text{entropy}} \), which encourage exploration. 
The overall loss is given by:

\begin{equation}
\resizebox{0.435\textwidth}{!}{$
L^{\text{PPO}}_t(\theta, \phi) = \mathbb{E}_t \left[ L^{\text{CLIP}}_t(\theta) - c_1 L^{\text{VF}}_t(\phi) + c_2 L^{\text{entropy}}_t(\theta) \right],
$}
\end{equation}
where \( c_1 \) and \( c_2 \) are  coefficients balancing the contributions of each term. 
The pseudocode for PPO is written in Algorithm \ref{PPO_AlG} \cite{lim2020federated}.  
\begin{algorithm}[!]
\caption{PPO (Actor-Critic Style)}
\begin{algorithmic}[1]
\STATE Initialize policy parameters \(\theta\) (Actor) and value function parameters \(\phi\) (Critic)
\FOR{each iteration}
    \STATE Collect trajectories using the current policy \(\pi_\theta\)
    \STATE Compute advantage estimates \(\hat{A}_t\) using Generalized Advantage Estimation (GAE)
    \STATE Optimize the surrogate objective \(L^{\text{PPO}}\) for a fixed number of epochs using gradient ascent
    \STATE Update policy parameters \(\theta\) and value function parameters \(\phi\)
\ENDFOR
\end{algorithmic}
\label{PPO_AlG}
\end{algorithm}


LIME is a model-agnostic interpretability method that explains individual predictions by approximating the behaviour of a complex model locally with a simpler, interpretable surrogate model \cite{LIME_ORIG}.
Given an input instance \(x\), LIME generates perturbed samples around \(x\), calculates the corresponding outputs using the original model, and fits a simple model \(g\) to these outputs. 
The optimization problem is:

\begin{equation}
\xi(x) = \argmin_{g \in G} \left( \mathcal{L}(f, g, \pi_x) + \Omega(g) \right),
\end{equation}
%
where \(\mathcal{L}(f, g, \pi_x)\) measures the fidelity of \(g\) in approximating the original model \(f\) locally, \(\pi_x\) is a proximity measure that assigns higher weights to samples closer to \(x\), and \(\Omega(g)\) is a complexity penalty ensuring the model has a low number of parameters.
The proximity function is defined as:

\begin{equation}
\pi_x(z) = e^{\left(-\frac{D(x, z)^2}{\sigma^2}\right)},
\end{equation}
where \(D(x, z)\) is the distance between \(x\) and a perturbed sample \(z\), and \(\sigma\) controls the locality.
LIME prioritizes samples closer to the instance \(x\), ensuring the explanation model \(g\) focuses on local behavior.


In this work, LIME is applied to interpret the decisions made by the actor in the PPO framework. 
Specifically, the actor's policy \(\pi_\theta(a_t | s_t)\), which determines the action probabilities given a state, is analysed locally using LIME.
By generating explanations for the actor's decisions, LIME helps uncover the factors influencing the policy's behavior. 
This interpretability is crucial for understanding and debugging reinforcement learning agents, especially in critical infrastructures like microgrid energy management.
By combining PPO’s robust policy optimization with LIME’s interpretability, this approach balances performance and transparency, enabling the DRL agent’s decisions to be both effective and explainable.

\section{Simulation Results}\label{section:Sim_Result}
\begin{figure}[h]
    \centering
    \includegraphics[width=0.48\textwidth]{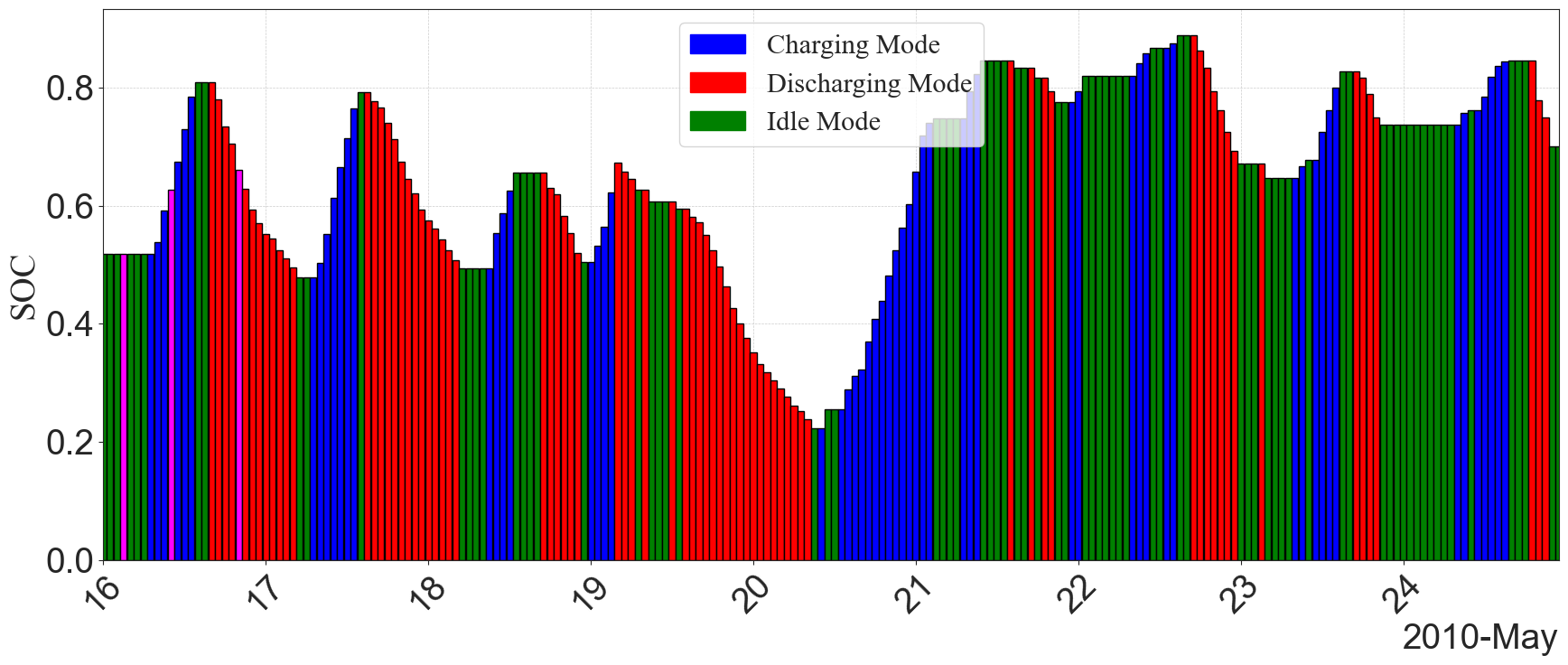}
    \caption{SOC plot during the cyclone.}
    \label{fig:SOC_DRL}
\end{figure}

\begin{figure}[h]
\centering
\subfloat[]{\includegraphics[width=0.48\textwidth]{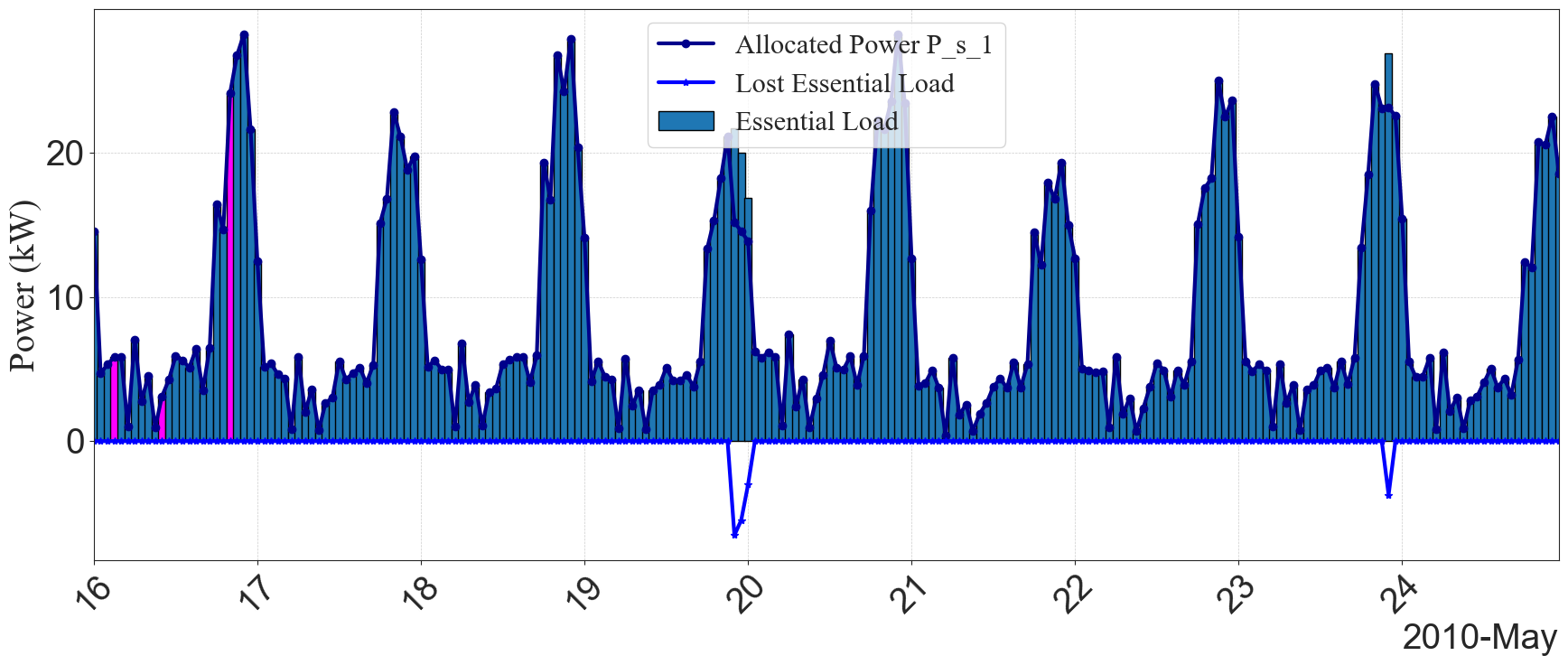} \label{fig:DRL_essential}}
\hfill
\subfloat[]{\includegraphics[width=0.48\textwidth]{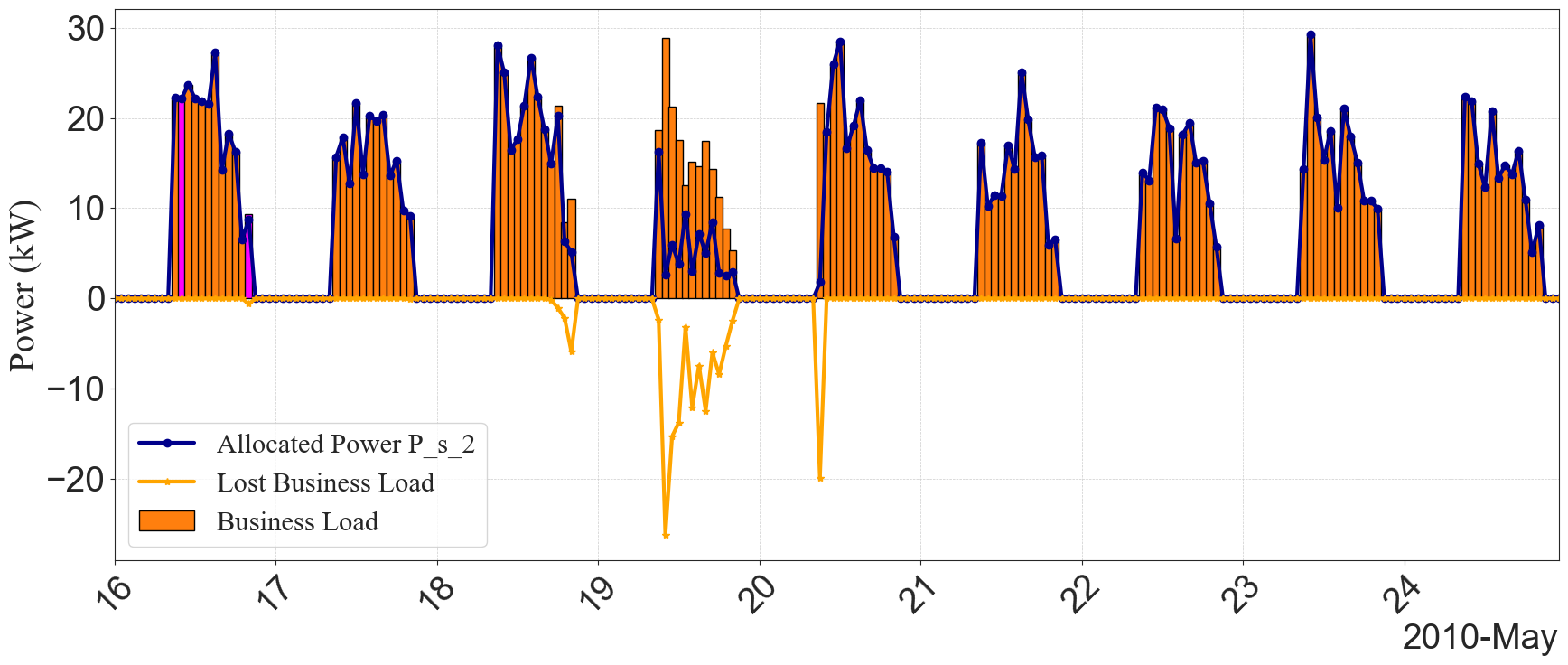} \label{fig:DRL_business}}
\hfill
\subfloat[]{\includegraphics[width=0.48\textwidth]{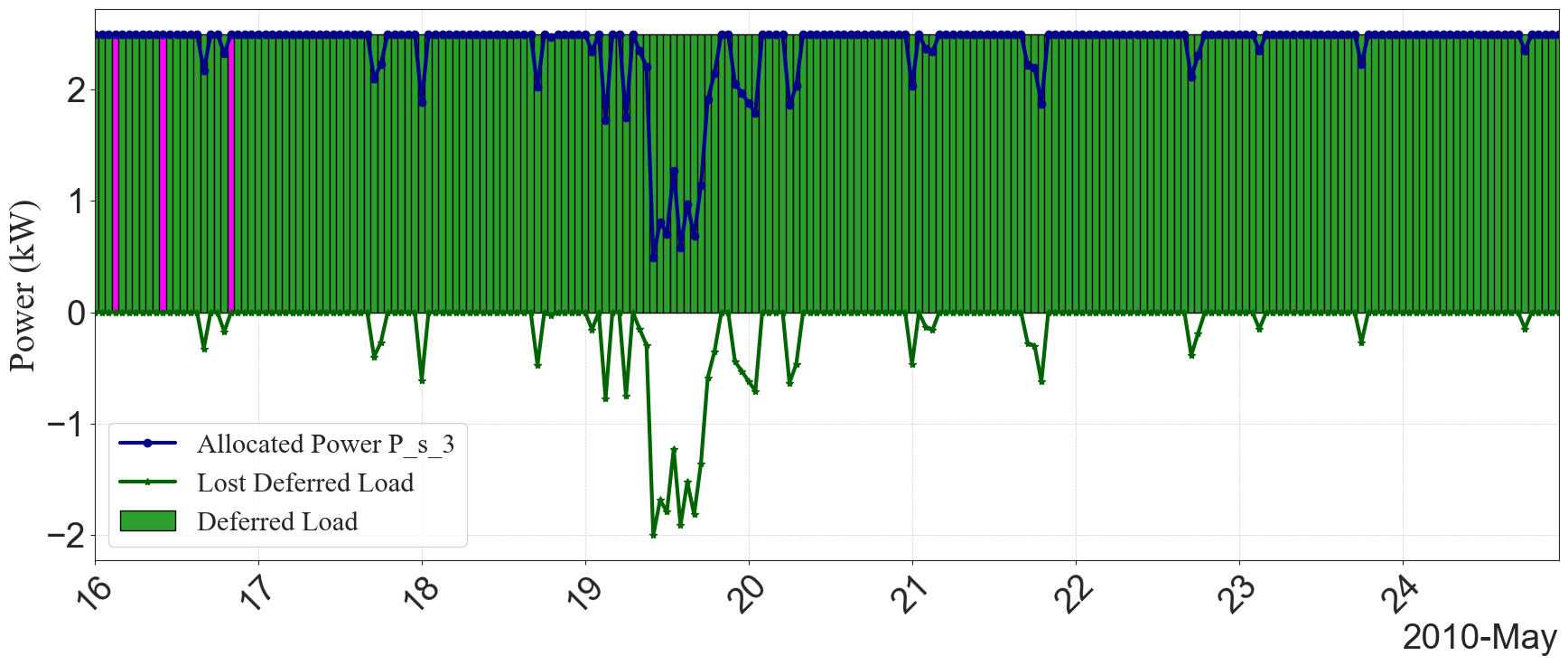} \label{fig:DRL_agriculture}}
\caption{a) Essential Load, b) Business Load, and c) Agricultural Load Demand, Supply, and Imbalances.}
\label{fig:Load1}
\end{figure}

This section presents the simulation results.
Figure \ref{fig:SOC_DRL} illustrates the battery's SOC, which indicates charging and discharging patterns over time. 
Figure \ref{fig:Load1} shows the load supply for the three prioritized loads.
The total resilience index achieved was 0.9736, a reasonable value for an event like a cyclone. 
\begin{figure}[h]
    \centering
    \includegraphics[width=0.48\textwidth]{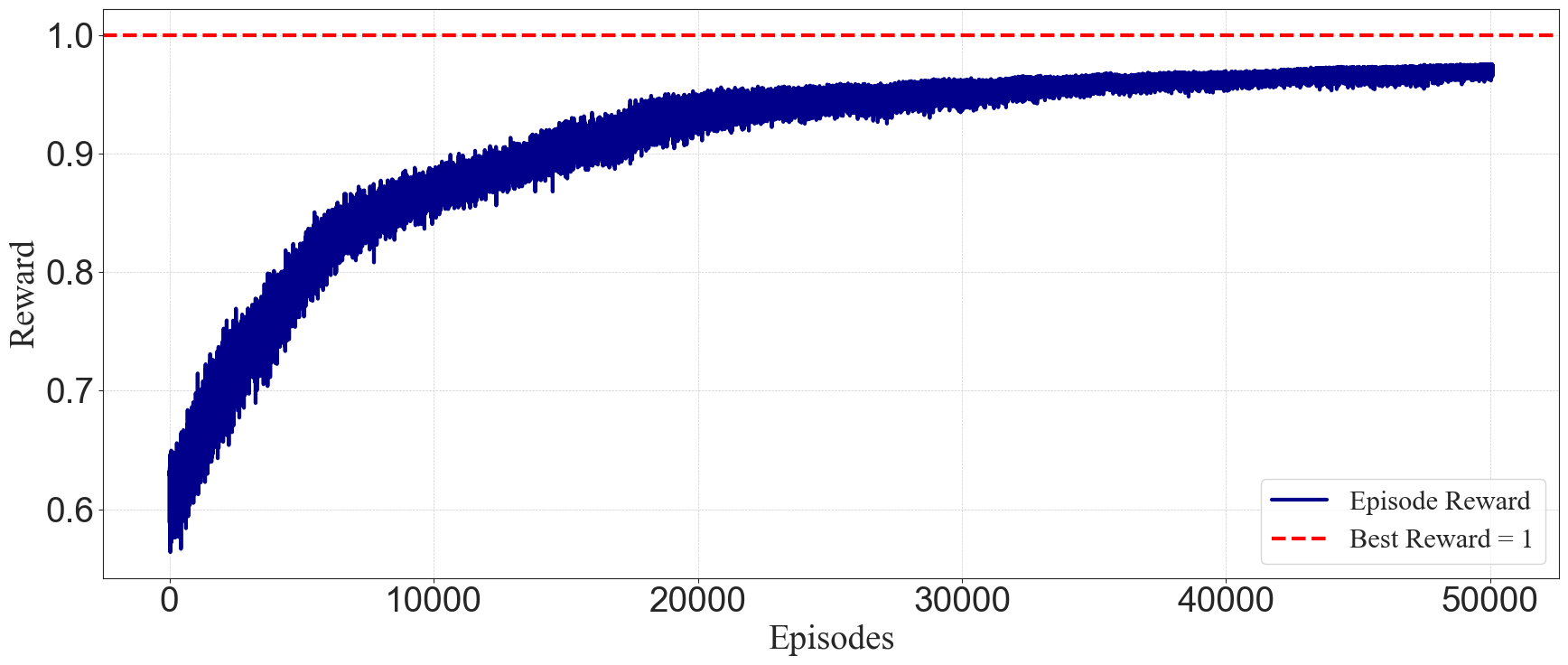}
    \caption{Reward Convergence Over episodes.}
    \label{fig:Reward}
\end{figure}
Additionally, the expected battery life was estimated to be 15.11 years, aligning with the design parameters from HOMER Pro.
Additionally, the reward convergence curve is shown in Figure \ref{fig:Reward}, where it can be observed that the reward stabilizes and converges after approximately 40,000 episodes and approaches the oracle strategy.
This article focuses on analysing the factors driving the DRL agent's decision-making process for battery charging and discharging.
In Figure \ref{fig:SOC_DRL}, three specific scenarios are shown in pink to represent distinct operational modes on day 16: idle at hour 4, charging at hour 11, and discharging at hour 21. 
The agent's choices are analysed in these scenarios to explore the factors influencing its decisions under different scenarios.

\begin{figure}[!]
\centering
\subfloat[]{\includegraphics[width=0.44\textwidth]{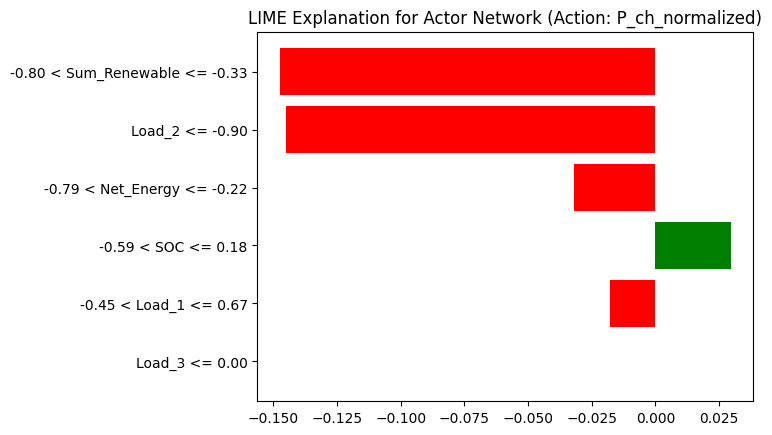} \label{fig:idle_CH}}
\hfill
\subfloat[]{\includegraphics[width=0.44\textwidth]{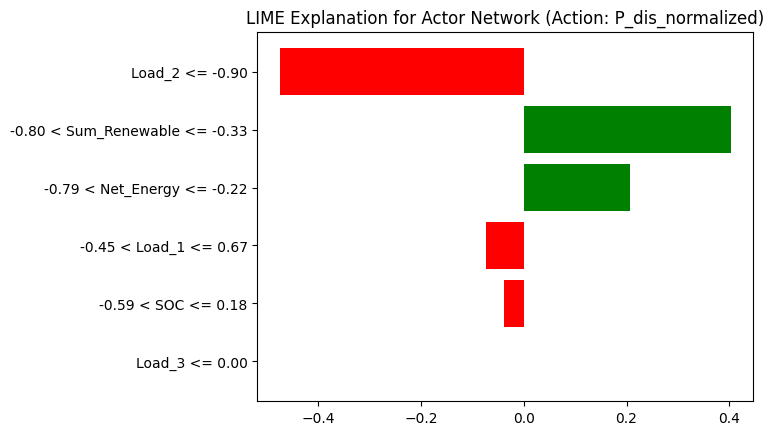} \label{fig:idle_DIS}}
\caption{LIME-based Explanations in IDLE Mode: a) Charging, and b) Discharging Action.}
\label{fig:IDLE_EXP}
\end{figure}

The LIME results for the idle mode, presented in Figures \ref{fig:idle_DIS} and \ref{fig:idle_CH}, indicate that the DRL actor's decision to neither charge nor discharge is driven by conflicting impacts of various features.
For the charging action (Figure \ref{fig:idle_CH}), all features discourage charging, with significant negative contributions from SOC, renewable generation, and Load\_2, suggesting an unfavourable condition for charging. 
In contrast, for the discharging action (Figure \ref{fig:idle_DIS}), SOC, renewable generation, and net energy positively influence discharging, but these are nearly counterbalanced by the high negative impact of Load\_2, resulting in no clear encouragement for discharging. 
This balance of influences leads the actor to select the idle action. 
Additionally, Load\_3 remains insignificant across all scenarios due to its fixed nature and low priority, while Load\_2 emerges as the most impactful feature, highlighting the actor's sensitivity to intermediate-priority loads in decision-making.

\begin{figure}[!]
\centering
\subfloat[]{\includegraphics[width=0.44\textwidth]{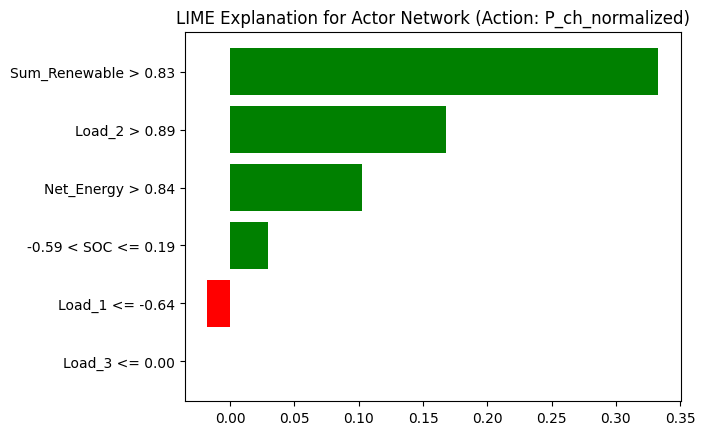} \label{fig:CH_CH}}
\hfill
\subfloat[]{\includegraphics[width=0.44\textwidth]{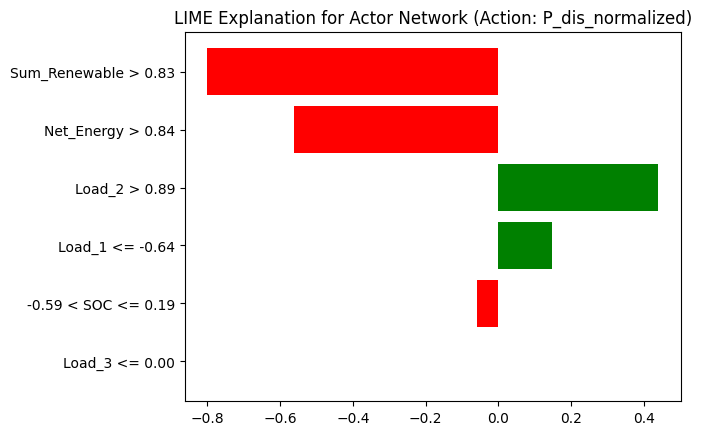} \label{fig:CH_DIS}}
\caption{LIME-based Explanations in Charging Mode: a) Charging, and b) Discharging Action.}
\label{fig:CH_EXP}
\end{figure}

In the charging scenario, illustrated in Figures \ref{fig:CH_CH} and \ref{fig:CH_DIS}, the actor’s decision to charge is strongly influenced by multiple encouraging features.
In the charging action (Figure \ref{fig:CH_CH}), SOC emerges as the most impactful positive factor, supported by contributions from renewable generation, Load\_2, and net energy, all favouring the charging action.
Conversely, in the discharging action (Figure \ref{fig:CH_DIS}), SOC, renewable generation, and net energy significantly discourage discharging, while Load\_2 and Load\_1 do not provide sufficient encouragement.
This leads the actor to select the charging action as the optimal choice in this instance.

\begin{figure}[!]
\centering
\subfloat[]{\includegraphics[width=0.44\textwidth]{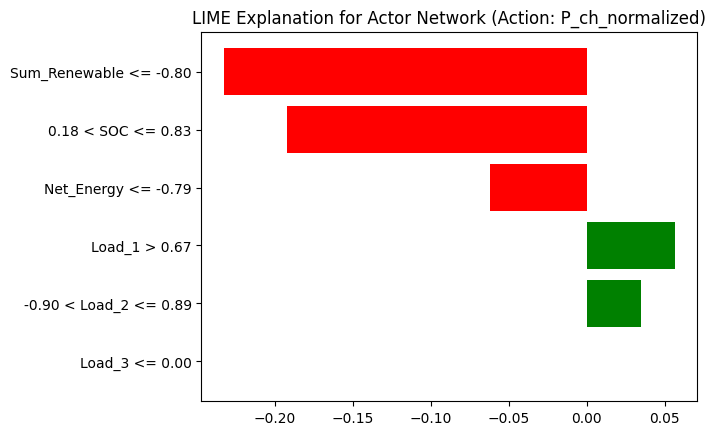} \label{fig:DIS_CH}}
\hfill
\subfloat[]{\includegraphics[width=0.44\textwidth]{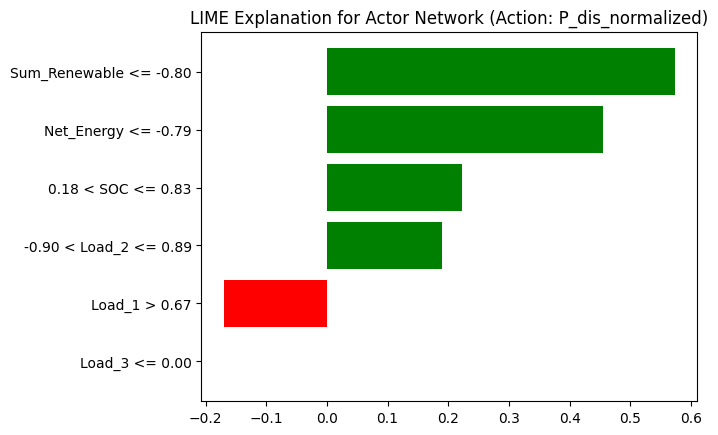} \label{fig:DIS_DIS}}
\caption{LIME-based Explanations in Discharging Mode: a) Charging, and b) Discharging Action. }
\label{fig:DIS_EXP}
\end{figure}

In the discharging scenario, shown in Figures \ref{fig:DIS_CH} and \ref{fig:DIS_DIS}, the actor’s decisions are shaped by contrasting feature impacts.
In the charging action (Figure \ref{fig:DIS_CH}), factors such as renewable generation, SOC, and net energy have significant negative impacts, discouraging the actor from charging.
However, in the discharging action (Figure \ref{fig:DIS_DIS}), these same features contribute positively, strongly encouraging discharging. 
This alignment of positive impacts on discharging explains the actor's choice to discharge in this scenario.

\section{Conclusion} \label{section:Conclusion}

This study introduced a transparent framework for microgrid resilience management by combining PPO and LIME. 
The microgrid was designed for the coastal city of Ongole in India, aligning with local load and weather conditions under the Layla cyclone scenario. 
The approach achieved a high Resilience Index (0.9736) while maintaining an estimated battery lifespan of over 15 years.
LIME identified key factors influencing decisions, including renewable generation, load priorities, and battery SOC, enhancing stakeholder confidence and trust in the agent's charging and discharging actions.

Future research can explore alternative explainability methods (e.g., SHAP) for a broader perspective and global explanations of the DRL agent’s decisions.
Investigations into multi-microgrid coordination, real-time constraints, and cybersecurity considerations would further strengthen overall resiliency. 
Additionally, incorporating economic factors and battery life into the reward function could balance resilience with profitability.

\bibliographystyle{ieeetr}
\bibliography{biblio}

\end{document}